# Femtosecond X-ray emission study of the spin cross-over dynamics in haem proteins

Dominik Kinschel[1], Camila Bacellar[1], Oliviero Cannelli[1], Boris Sorokin[1], Tetsuo Katayama[2], Giulia F. Mancini[1], Jeremy R. Rouxel[1], Yuki Obara[3], Junichi Nishitani[4], Hironori Ito[3], Terumasa Ito[3], Naoya Kurahashi[5], Chika Higashimura[4], Shotaro Kudo[4], Theo Keane[6], Frederico A. Lima[7], Wojciech Gawelda[7,8], Peter Zalden[7], Sebastian Schulz[7], James Budarz[1], Dmitry Khakhulin[7], Andreas Galler[7], Christian Bressler[7], Christopher J. Milne[9], Thomas Penfold[6], Makina Yabashi[2], Toshinori Suzuki[4], Kazuhiko Misawa[3] and Majed Chergui[1]*

[1]*Laboratoire de Spectroscopie Ultrarapide (LSU) and Lausanne Centre for Ultrafast Science (LACUS), Ecole Polytechnique Fédérale de Lausanne (EPFL), 1015 Lausanne, Switzerland*
[2]*Japan Synchrotron Radiation Research Institute (JASRI), 1-1-1, Kouto, Sayo-cho Sayo-gun, Hyogo, 679-5198, Japan*
[3]*Tokyo University of Agriculture and Technology (TUAT), 2-24-16 Naka-cho, Koganei, Tokyo 184-8588 Japan*
[4]*Department of Chemistry, Graduate School of Science, Kyoto University, Kitashirakawa-Oiwakecho, Sakyo-Ku, Kyoto 606-8502, Japan*
[5]*Department of Materials and Life Sciences, Faculty of Science and Technology, Sofia University, 102-8554 Tokyo, Chiyoda, Kioicho, 7-1, Japan*
[6]*Chemistry-School of Natural and Environmental Science, Newcastle University, Newcastle upon Tyne, NE1 7RU United Kingdom*
[7]*European XFEL, Holzkoppel 4, 22869 Schenefeld, Germany*
[8]*Faculty of Physics, Adam Mickiewicz University, 61-614 Poznań, Poland*
[9]*SwissFEL, Paul-Scherrer-Institut (PSI), 5232 Villigen, Switzerland*

## Abstract

In haemoglobin (consisting of four globular myoglobin-like subunits), the change from the low-spin (LS) hexacoordinated haem to the high spin (HS) pentacoordinated domed form upon ligand detachment and the reverse process upon ligand binding, represent the "transition states" that ultimately drive the respiratory function. Visible-ultraviolet light has long been used to mimic the ligand release from the haem by photodissociation, while its recombination was monitored using time-resolved infrared to ultraviolet spectroscopic tools. However, these are neither element- nor spin-sensitive. Here we investigate the "transition state" in the case of Myoglobin-NO (MbNO) using femtosecond Fe $K_\alpha$ and $K_\beta$ non-resonant X-ray emission spectroscopy (XES) at an X-ray free-electron laser upon photolysis of the Fe-NO bond. We find that the photoinduced change from the LS ($S = 1/2$) MbNO to the HS ($S = 2$) deoxy-myoglobin (deoxyMb) haem occurs in ~800 fs, and that it proceeds via an intermediate ($S = 1$) spin state. The XES observables also show that upon NO recombination to deoxyMb, the return to the planar MbNO ground state is an electronic relaxation from HS to LS taking place in ∼30 ps. Thus, the entire ligand dissociation-recombination cycle in MbNO is a spin cross-over followed by a reverse spin cross-over process.

Small ligand molecules ($O_2$, CO, NO, CN and $H_2O$) are the receptors that bind to and activate haem proteins, such as myoglobin (Mb) and haemoglobin (Hb). Their active site (Figure 1) is a porphyrin with an Iron (that can be either ferric-$Fe^{3+}$ or ferrous-$Fe^{2+}$) at its centre to which the ligands bind on the distal side. The haem itself is linked to the F helix of the protein via a histidine (His) amino-acid bound to the Fe ion on the proximal side of the porphyrin. The ferrous forms of Mb and Hb play an essential role in respiration via the transport and storage of oxygen in mammals. In the ligated form, the hexacoordinated $Fe^{2+}$ ion of the haem is in a low-spin (LS) planar singlet (MbCO) or doublet state (MbNO). Upon release of the ligand from the $Fe^{2+}$ ion, the haem switches to a high-spin (HS) quintet state with the pentacoordinated Fe in an out-of-plane, so-called domed, deoxyMb configuration (Figure S1). The shuttling between the LS planar to HS domed spin cross-over (SCO) process that defines the "transition state", which governs the allosteric transitions in haem proteins (including $O_2$ carriers and NO- and CO-sensors).[1,2] In particular it is the first event of the respiratory function in which, after release of the distal ligand, the Fe atoms moves out of the porphyrin plane, pushing the proximal histidine (His) linked to the F helix and displacing the latter (Figure 1). This is considered the step by which the monomer units of Hb transfer information, ultimately leading to the transition from the so-called "relaxed" to "tense" state at later times.[3]

It has long been known that ligand detachment in ferrous Mb's can be induced by visible-ultraviolet (UV) excitation of π−π* transitions of the haem.[4] This has been exploited to mimic the release of the distal ligand from the Fe centre, and monitor its recombination to the haem. With the advent of ultrafast spectroscopy, haem proteins were among the first systems ever to be studied, specifically with the aim of monitoring these processes in "real-time" using transient absorption (TA) spectroscopy from the visible-UV (sensitive to the porphyrin π−π* transitions)[5-9] to the Infrared (sensitive to the ligand stretch and the porphyrin vibrational modes)[10,11] as well as by time-resolved resonance Raman spectroscopy (sensitive to the Fe-His bond).[12,13] These studies concluded that photodissociation of the diatomic ligands and doming are prompt and simultaneous events, typically in <50-100 femtoseconds (fs).[14] The actual formation of the ground state HS deoxyMb form has been described by two main and non-mutually excluding models: a) a vibrationally hot ground-state is formed promptly after excitation, in which cooling ensues;[8,15,16] b) the process is a cascade through intermediate electronic excited states.[6,17]

Recently, the occurrence of prompt doming was questioned in a fs hard X-ray absorption spectroscopy (XAS) study of photoexcited MbCO.[18] Time-resolved XAS is an ideal element-selective probe of the electronic and geometric structure changes of metal-containing molecular complexes.[19] Using a monochromatic probe at 7.123 keV (tuned to the Fe K-edge), Levantino et al[18] observed a first <50–70 fs event, which they attributed to CO photolysis from the LS (S=0) MbCO and partial SCO to a higher spin state, followed by a further 300-400 fs event, attributed to the passage to the HS (S=2) state of deoxyMb. This picture was supported by quantum wave packet dynamics simulations,[20] which accounted for coherent nuclear and electronic motion. However, Levantino *et al*'s[18] use of a monochromatic probe hindered a spectroscopic identification of the intermediate and final states, let alone of their spin. In recent years, non-resonant X-ray emission spectroscopy (XES) has emerged as a valuable tool to identify spin states of transition metal complexes (TMC).[21] The $K_{\beta 1,3}$ and $K_{\beta'}$ lines ($3p \rightarrow 1s$ emission, Fig. 2a) have been used as markers of the spin state (number of unpaired $3d$ electrons) of TMC's via the $3p$-$3d$ exchange interactions and fs transient $K_\beta$ XES was successfully implemented to monitor the SCO dynamics in $[Fe(bpy)_3]^{2+}$,[22] as well as to identify the HS deoxy haem product of photoexcited ferrous Cytochrome c after dissociation of its distal methionine ligand.[23] In the case of 2p-orbitals, the p-d exchange is weaker but $K_\alpha$ XES ($2p \rightarrow 1s$ transitions, Fig. 2a) can also be used as a marker of spin state via the linear dependence of the full-width at half-maximum (FWHM) of the $K_{\alpha 1}$ line as a function of the number of unpaired metal $3d$ electrons.[24]

In this contribution, we focus on the case of Nitrosylmyoglobin (MbNO). The role of nitric oxide (NO) in physiological processes in humans was discovered in the 1990s.[25] In its interaction with haem systems, the NO molecule plays an important role in several physiological effects and biological functions such as neurotransmission,[26] regulation of vasodilatation,[27] platelet aggregation,[28] and immune response.[29] Furthermore, studies of NO release from and binding to the haem provide mechanistic information of relevance for other ligands active in respiration, blood pressure regulation, nitrogen fixation, and NO oxidation. In this respect, several time-resolved studies using visible-ultraviolet and infrared (IR) probes have been carried out on MbNO over the past three decades. Just as for MbCO, NO photodissociation from the haem has been reported to be prompt (<100 fs) with a quantum yield of ~55 %,[7] and recovery of the system occurs on typical timescales of 1-3 ps, 5-20 ps, 110-290 ps, and a weak (~10-20%) ns component.[7,11,13,30-37] The shortest recovery time was attributed to cooling within what was considered, the promptly formed hot ground state of deoxyMb,[8] while the next two decay times were attributed to geminate recombination to the latter of, respectively, NO ligands from the distal pocket close to the haem and NO ligands from the more distant Xe4 pocket. Finally, the nanosecond (ns) component is due to non-geminate recombination of a small fraction of NO ligands that have escaped the protein. Contrary to other ligands (CO in particular), which can bind to a planar haem, the unpaired NO electron allows its binding to the domed HS deoxyMb,[36] and this was recently supported by a time-resolved Fe K-edge X-ray near edge structure (XANES) study with 70 ps resolution.[37] The binding of NO to a domed deoxyMb (hereafter designated as deoxyMb-NO) had previously been predicted by theoretical calculations on model haems,[38] which argued that the return to the planar form is a reverse SCO from HS deoxyMb-NO to LS planar MbNO. Recent ultrafast resonance Raman studies[13] monitoring the Fe-His bond (Figure 1) determined that the domed-to-planar transition in MbNO takes place in ~30 ps and argued that the process is due to constraints exerted by the protein structure on the haem cofactor (Figure 1). Since the entire ligand dissociation-recombination cycle occurs within ~250 ps, and is much shorter than the commonly studied MbCO, this implies that it can be monitored within the same ultrafast spectroscopic experiment.

In order to identify the mechanisms leading to the formation of HS deoxyMb in Myoglobins and its time scale, and the specific case of the reverse deoxyMb to MbNO transition, here we use fs Fe $K_\alpha$ and $K_\beta$ non-resonant X-ray emission spectroscopy (XES) with a von Hamos spectrometer (Figure 2b) at the SPring-8 Angstrom Compact Free Electron Laser (SACLA).[39,40] We recorded laser-off and laser-on XES spectra (whose difference yields the transient XES) as a function of time delay with respect to the pump pulse. We find that upon 533 nm photoexcitation of MbNO into the Q-band (Figure S2), the SCO from the ground state LS ($S=1/2$) MbNO[41] to the HS ($S=2$) deoxyMb occurs in ~800 fs and involves the passage via an intermediate $S=1$ spin state, while the return from the HS deoxyMb-NO to the LS MbNO is a reverse SCO that takes place in ~30 ps. We thus establish for the first time the details of domed HS deoxyMb formation, which we argue is of general validity to all Mb's, and we identify the return to the initial MbNO ground state also as a HS-LS relaxation. The observation of the $S=1$ state is supported by density functional theory (DFT) calculations of the $K_\alpha$ XES. Details of the experimental set-up and procedures and of the theoretical calculations are given in the SI.

In previous ultrafast[22,23] or quasi-static[42] $K_\beta$ XES studies, the assignment of spin states relied on the comparison with reference spectra of model compounds. Figure S3 shows the steady-state $K_\beta$ emission spectra of hexacoordinated MbNO and pentacoordinated deoxyMb and compares them with those of reference Fe-based molecular complexes in figure S3b,[22] and reference Fe(II) LS and HS porphyrin compounds.[42] While the trends are similar as a function of increasing the spin in all three panels, with the intensity increase of the $K_{\beta'}$ line around 7045 eV and a blue shift and intensity decrease of the $K_{\beta1,3}$ line near 7058 eV, some differences also show up in that the blue shift of the $K_{\beta1,3}$ line in figure S3a and its energy splitting with respect to the $K_{\beta'}$ line are smaller than in figures S3b and c. These differences may point to a covalency change[43-46] in addition to the spin change, as will be discussed later. Figure S4a shows the steady-state $K_\alpha$ lines of MbNO and deoxyMb. There is an intensity decrease and slight blue shift and broadening of the $K_{\alpha1}$ line with increasing spin, as reported for other compounds,[24] while the $K_{\alpha2}$ line undergoes an intensity decrease and a slight broadening. Because $K_\alpha$ XES spectra of reference compounds are lacking, we resorted to *density functional theory* (DFT) to simulate them (see § S4), as discussed later. In summary, the above XES spectra for the Mb LS hexacoordinated and the HS pentacoordinated forms provide the two extreme cases before and after photodissociation of the ligand, expected in time-resolved experiments, which allow us to identify the intermediate steps.

The laser-off and laser-on $K_\beta$ spectra are shown in figure 3 for time delays up to ~1.4 ps, while the inset zooms into the region of the maximum of the $K_{\beta1,3}$ line. These spectra show an intensity decrease and blue shift of the latter, an intensity decrease in the 7050-7055 eV region and a slight increase of the $K_{\beta'}$ line in the 7045-7050 eV region, all of which reflect the trends predicted for a change from LS to HS Mb (Figure S3a) and qualitatively agree with reference Fe(II) complexes (Figure S3b and c).[22] Laser-on spectra at later times are shown in Figure S5, and they exhibit a return to the laser-off spectra on tens to hundreds of ps. Further insight into these changes is obtained from the normalised transient $K_\beta$ XES spectra (excited minus ground state signal) shown in Figure S6a for time delays up to 1.36 ps, and Figure S6b for time delays up to 100 ps. These transients exhibit a derivative-like shape near 7058 keV, a negative signal that extends down to 7050 eV and a positive one below in the region of the $K_{\beta'}$ line. Beyond ~1 ps, the normalized transient spectra no longer change but before, the 7050-7057 eV region shows a gradual increase of negative amplitude, while the peak near 7060-7062 eV is most blue shifted at 0.26 ps, and then shifts slightly to the red and stabilizes at 7060 eV for times >1 ps. The evolution of the 7050-7057 eV region qualitatively reproduces the static difference spectra of reference Fe-complexes of different spins (Figure S7), that revealed the transient population of a triplet state in the photoinduced SCO of $[Fe(bpy)_3]^{2+}$.[22] From figures 3 and S6a, we conclude that reaching the final HS deoxyMb state requires ~1 ps and proceeds via an intermediate state.

Figure 4 shows the laser-off and laser-on $K_{\alpha 1}$ and $K_{\alpha 2}$ XES lines at different time delays, while the insets zoom into the region of their maxima. One notices a clear gradual weakening and a slight broadening (and for $K_{\alpha 1}$, a slight blue shift) of the bands within the first ps. The trends in energy, width and intensity of these lines are consistent with the differences we observe between MbNO and deoxyMb (Figure S4), but in Figure 4, the evolution is gradual and on a similar time scale as for the $K_\beta$ lines. Figure S8 shows the laser-on spectra at later times, and just as in figure S6, they exhibit a return to the laser-off spectra on tens to hundreds of ps. In order to highlight the changes, figure S9 shows the experimental $K_\alpha$ XES transients at 0.26 and 1.36 ps time delay, while those at intermediate and at later times are shown in Figure S10. The normalized transient line shapes reflect the broadening, shift and intensity changes of the asymmetric emission lines, which do not change beyond ∼1 ps, by which time the system is in the HS state, as confirmed by the difference of the experimental steady-state spectra deoxyMb minus MbNO (Figure S9b). In summary, both the $K_\alpha$ and $K_\beta$ XES transients point to an intermediate state being populated prior to the HS deoxyMb state that is fully formed by ~1 ps.

The signal around 7053 eV in Figures 3 and S6 best differentiates the intermediate state from the initial excited LUMO state to the quintet state. This is where the signature of an intermediate spin state was most pronounced in the previous study on $[Fe(bpy)_3]^{2+}$.[22] By plotting its amplitude as a function of time for the transients available up to 3.6 ps, we can get an estimate of the lifetime of the intermediate state, using a 3-level kinetic model (§ S10 and figure S11), assuming 100% Q-state population at t=0 that decays to the triplet state in ~100 fs, as concluded from ultrafast fluorescence up-conversion studies.[47] The fit yields a lifetime of 500 ± 250 fs for the intermediate state (Figures S12 and S13).

Further details into the ultrafast kinetics comes from the peak position of the laser-on $K_{\beta_{1,3}}$ line (Figures 3 and S5) and the full width at half maximum (FWHM) of the laser-on $K_{\alpha 1}$ line (Figures 4 and S8), both of which are sensitive to spin. During the first ps, these spectra reflect a mixture of doublet, short-lived intermediate state and of the final quintet state, before the entire excited population settles in the latter. The $K_{\beta 1,3}$ peak energy shift and the FWHM of the $K_{\alpha 1}$ line are plotted in Figure 5 for times up to 750 ps, while the inset zooms into the first 10 ps. The red trace is a fit of the experimental data using a function consisting of a rising component and a biexponential decay. The former corresponds to a risetime of 800 ± 150 fs (see § S6), significantly longer than our cross-correlation (∼150 fs), while the decay components have time constants (pre-exponential factors) of $\tau_1 = 30 \pm 9$ ps ($A_1=0.6$) and $\tau_2= 1.5 \pm 0.9$ ns ($A_2=0.4$). Interestingly, while the UV-visible and IR TA results are characterized by a prompt risetime and decay times of 5-20 ps, 110-290 ps, and a weak ns component,[7,11,13,30-37] the present XES observables exhibit different kinetics with a rise of ~800 fs and decays of ~30 ps and ns's. As mentioned previously, the ~30 ps decay had been reported in the time-resolved resonance Raman studies by Kruglik et al,[13] who attributed it to the relaxation from the domed HS deoxyMb-NO species to the LS planar MbNO. Given the spin sensitivity of XES our observation of the ~30 ps decay is fully consistent with this interpretation as it reflects the return of the system to the LS ground state. Since the trigger to this process is the binding of the NO ligand to deoxyMb, in principle the geminate recombination (in 5-20 ps and 110-250 ps) and the non-geminate recombination (ns component) should feed the 30 ps component. In this scenario, the 5-20 ps component would appear as a rise time, while the 110-250 ps and the ns components would show up as decay because they are rate-determining to the 30 ps component. However, due to the scatter of data points at early times in figure 5, we refrained from extracting such times from the fits, preferring the above phenomenological function, which captures the important features using a single rise time and a biexponential decay. In this respect, the relatively large pre-exponential factor of the ns decay component, compared to previous reports (see [37] and references therein), most likely reflects the fact that it includes both geminate (110-250 ps) and non-geminate (ns) recombination of NO ligands.

Coming back to the short time dynamics, we have yet to identify the intermediate state between the initial MbNO excited state and the HS deoxyMb. The changes occurring at early times (Figure 3 and S5) reflect the transient population of an intermediate, most likely an S=1 spin state, if we parallel the present results with those on photoexcited [Fe(bpy)$_3$]$^{2+}$.[22] We further support the assignment of the intermediate state as having S=1 using DFT calculations, presented in § S4 and Figures S14 and S15. However, in addition to its spin sensitivity, K$_\beta$ XES is also sensitive to metal-ligand covalency,[43-46] which represents the charge donation from the ligands to the metal and therefore, it has a significant influence on the nature of metal-ligand bonds. The degree of covalency is connected via the effective number of unpaired 3d electrons to the magnitude of the exchange splitting. In K$_\beta$ XES, it manifests itself in, among others, the splitting between the K$_{\beta1,3}$ and K$_{\beta'}$ lines. Thus, covalency could also affect the transient spectra as a result of dissociating the NO ligand and of the relaxation to the HS state, which involves an elongation of the Fe-Np bonds (Np are the pyrrole nitrogens). We hold however unlikely that covalency effects affect our results: a) Given the recombination times of NO to deoxyMb (5-20 ps and 110-290 ps), it seems surprising that the $K_\beta$ transient XES no longer changes beyond 1 ps (Figure S5). This is consistent with the $K_\beta$ XES predominantly reporting on the spin state of the metal, i.e. since it does not distinguish between the HS deoxyMb and HS deoxyMb-NO species; b) Also, the invariance of spectra in the relaxation from NO-deoxyMb to planar MbNO in ~30 ps is to note, as this corresponds to a contraction of the Fe-Np bonds; c) Thus, neither NO rebinding nor the ensuing relaxation from domed to planar seem to affect the $K_\beta$ XES. A fortiori, we would also exclude a contribution of covalency at early times (< 1 ps), since the reverse processes occur therein (ligand detachment, Fe-Np bond elongation); d) It should also be noted that the $K_\beta$ XES changes reported in the case of photoexcited [Fe(bpy)$_3$]$^{2+}$ where a significant bond elongation occurs[48] towards the HS state, were fully interpreted in terms of transient population via an intermediate spin state based on reference spectra;[22] f) Last but not least, covalency effects have not been reported on $K_\alpha$ XES and the very close correspondence between the temporal evolutions of the $K_\alpha$ and $K_\beta$ XES (Figure 5) suggests that the signature of covalency is negligible in the latter.

To summarize our results, Figure 6 shows the complete photocycle of MbNO: Upon photoexcitation into the Q-bands, prompt (<100 fs) photodissociation of NO occurs leaving an excited pentacoordinated haem that undergoes doming by first relaxing to an intermediate spin state (S=1), followed by a second relaxation step to the HS S=2 state forming deoxyMb. The entire cascade occurs in ~800 fs. The relaxation back from the domed HS haem to the planar ground state occurs in ~30 ps upon NO recombination. The latter takes place over several time scales due to geminate (5-20 ps, 110-200 ps) and non-geminate recombination (ns's). These results also show that the relaxation from domed NO-deoxyMb to the planar MbNO is a genuine HS to LS relaxation, as theoretically predicted by Franzen,[38] although constraints due to the F-helix (Figure 1)[13] cannot be fully excluded. Thus, the entire cycle of detachment and rebinding of NO to deoxyMb is accounted for by a SCO followed by a reverse SCO (Figure 6). There are many parameters that affect the rates of forward- and back-SCO but in the context of our interpretation, it is clear that the former is a stepwise process by jumps of $\Delta S=1$, which are ultrafast in metal complexes,[49] while the reverse SCO is a $\Delta S=2$ process occurring on a much slower time scale.

The above results have several implications on the description of the energy flow in haem proteins: a) upon ligand detachment, the formation of the fully domed deoxyMb requires ~800 fs, as it proceeds via an intermediate triplet state, which decays in 500 ± 250 fs; b) The ligand detachment is here mimicked by a photodissociation, which is prompt (<50 fs), implying that the ensuing dynamics is taking place in the pentacoordinated haem and should therefore be the same for all ferrous haems involved in ligand release. This was already suggested by Ye et al[8] based on UV-visible TA studies, but they concluded that the ensuing evolution within the deoxyMb product is due to relaxation of a vibrationally hot species, while we conclude that it is an electronic relaxation cascade. The debate to determine the initial pathways for ultrafast energy flow in haem proteins, has been going one for over 30 years now, with the interpretation about formation of deoxyMb going from an electronic cascade via spin states,[5,6] to a vibrational energy redistribution.[8,16] Here by using element- and spin-specific observables (XES), that are not sensitive to thermal and vibrational effects, we show that the initial scenario of a cascade among spin states is operative. Nevertheless, the cascade is non-radiative and energy gaps between spin states are dissipated in the form of heat to the environment. We believe that this behaviour is not limited to ferrous haems (that undergo distal ligand dissociation) but it also includes ferric haems (that remain hexacoordinated) as suggested by the conclusions of UV-visible studies on metMb[50] and MbCN.[51] Finally, doming has been suggested to trigger large scale conformational changes of the protein, the so-called "protein quake", that opens a channel for ligand escape from the protein.[52] Our results show this is unlikely to be a very fast process, contrary to earlier claims that assumed it to occur on the time scale of the Fe-ligand bond breaking.[53]

The present works identifies the transition from the planar ligated haem involved in ligand release to the domed deoxy form as a spin cross-over, which we believe is identical in all ferrous haem proteins. However, the ensuing events following ligand recombination are specific to the NO ligand and are characterized by a reverse SCO. This work also shows the power of ultrafast X-ray emission spectroscopy at unravelling exquisite details of the haem transformations upon ligand release and uptake in haem proteins.

**Methods Summary**

Femtosecond hard X-ray emission spectroscopy (XES) measurements were carried out on 4 mM solutions of MbNO in a physiological medium (pH = 7) under inert conditions (He or $N_2$ atmosphere) at BL3 at SACLA.[40] The sample solution was delivered through a 0.2 mm-thick round liquid jet and its integrity was continuously monitored by a mobile UV-VIS spectrometer. A laser pulse at 533 nm with a ~45 fs full width at half maximum (FWHM) was used to excite the Q-bands of MbNO in a near colinear geometry with the X-ray beam. An energy dispersive X-ray emission spectrometer (von Hamos geometry) and a 2D MPCCD detector are used to record the iron 2p-1s ($K_\alpha$) and 3p-1s ($K_\beta$) fluorescence spectra. A timing tool was used to measure the X-ray/optical relative arrival time fluctuations on a pulse to pulse basis and sort each shot by its relative arrival time (σ = 150 ± 40). For the analysis of the $K_\beta$ transients, we have used reference spectra of Fe-containing compounds from ref. [22]. For the $K_\alpha$ emission, reference spectra are lacking, and we relied on Density Functional Theory (DFT) simulated XES spectra for the ground state of MbNO and states of different spins for deoxyMb (See § S4).

**Acknowledgement**

This work was supported by the Swiss NSF via the NCCR:MUST and grants 200020_169914 and 200021_175649 and the European Research Council Advanced Grant H2020 ERCEA 695197 DYNAMOX. CB and GFM benefited from the InterMUST Women Postdoc Fellowship. The backing of the technical staff at SACLA is highly acknowledged. The experiments were performed with the approval of the Japan Synchrotron Radiation Research Institute (JASRI; Proposal No. 2017B8048). TP acknowledge support from the Leverhulme Trust (RPG-2016-103). W.G. acknowledges financial support by European XFEL and partial financial support from the Polish National Science Centre (NCN) under SONATA BIS 6 Grant No. 2016/22/E/ST4/00543.

**Author Information**

*Laboratoire de Spectroscopie Ultrarapide (LSU) and Lausanne Centre for Ultrafast Science (LACUS), Ecole Polytechnique Fédérale de Lausanne (EPFL), 1015 Lausanne, Switzerland*
Dominik Kinschel, Camila Bacellar, Oliviero Cannelli, Boris Sorokin, Giulia F. Mancini, Jérémy Rouxel, James Budarz, Majed Chergui

*Japan Synchrotron Radiation Research Institute (JASRI), 1-1-1, Kouto, Sayo-cho Sayo-gun, Hyogo, 679-5198, Japan*
Tetsuo Katayama

*Tokyo University of Agriculture and Technology (TUAT), 2-24-16 Naka-cho, Koganei, Tokyo 184-8588 Japan*
Yuki Obara, Hironori Ito, Terumasa Ito, Kazuhiko Misawa

*Department of Chemistry, Graduate School of Science, Kyoto University, Kitashirakawa-Oiwakecho, Sakyo-Ku, Kyoto 606-8502, Japan*
Junichi Nishitani, Chika Higashimura, Shotaro Kudo, Toshinori Suzuki

*Department of Materials and Life Sciences, Faculty of Science and Technology, Sofia University, 102-8554 Tokyo, Chiyoda, Kioicho, 7-1, Japan*

Naoya Kurahashi

*European XFEL, Holzkoppel 4, 22869 Schenefeld, Germany*
Frederico A. Lima, Wojciech Gawelda, Peter Zalden, Sebastian Schulz, Dmitry Khakhulin, Andreas Galler, Christian Bressler

*Faculty of Physics, Adam Mickiewicz University, 61-614 Poznań, Poland*
Wojciech Gawelda

*Chemistry-School of Natural and Environmental Science, Newcastle University, Newcastle upon Tyne, NE1 7RU, England*
Thomas Penfold, Theo Keane

*SwissFEL, Paul-Scherrer-Institut (PSI), 5232 Villigen, Switzerland*
Christopher J. Milne

## Contributions

MC conceived the project. MC, DK, CB, and OC designed the experiment. DK, CB, OC, BS, TK, YO, HI, TI, JN, CH, SK, MY and NK performed the experiment at SACLA. DK, CB, OC, BS, GFM, JB, FL, WG, PZ, SS, AG, DKh recorded the steady-state XES at Eu-XFEL. DK, CB, JR and MC analyzed the data. TP and TK performed the DFT simulations. DK and MC wrote the manuscript. All authors discussed the results and contributed to the manuscript.

## Competing interests

The authors declare no competing financial interests.

## Corresponding author

Correspondence to [Majed.Chergui@epfl.ch](Majed.Chergui@epfl.ch)

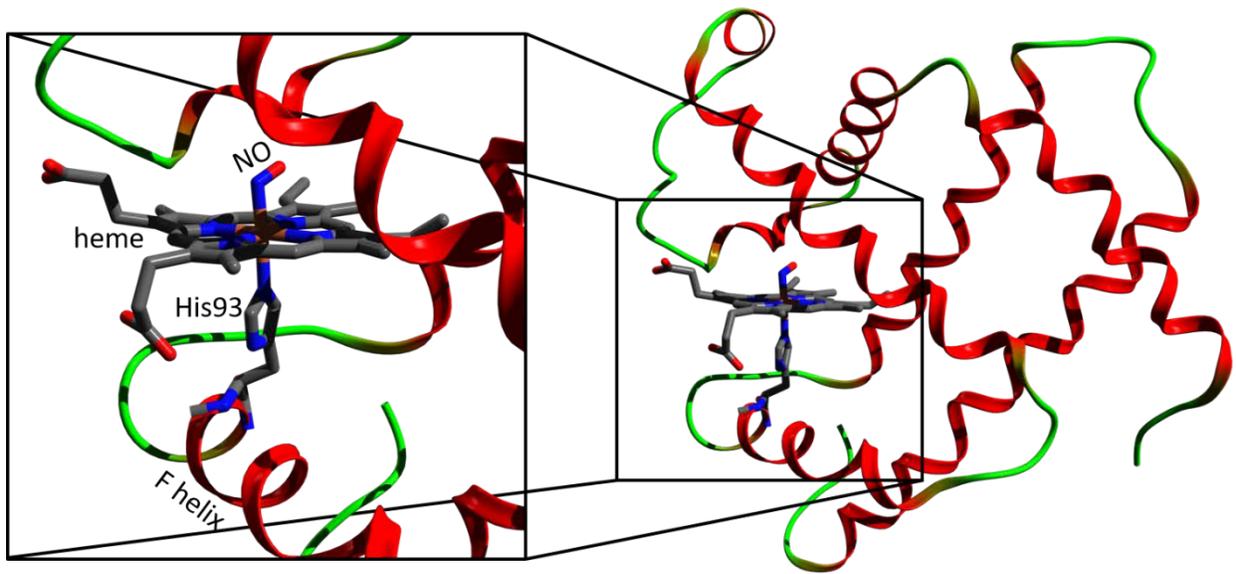

Figure 1: Crystal structure of the Nitrosyl-Myoglobin (MbNO). The haem is highlighted as sticks (Fe [orange], C [grey], N [blue], O [red]). PDB: 1HJT. The zoom (left) shows the distal NO ligand and proximal histidine 93 (His93), which links the Fe atom of the haem porphyrin to the F helix.

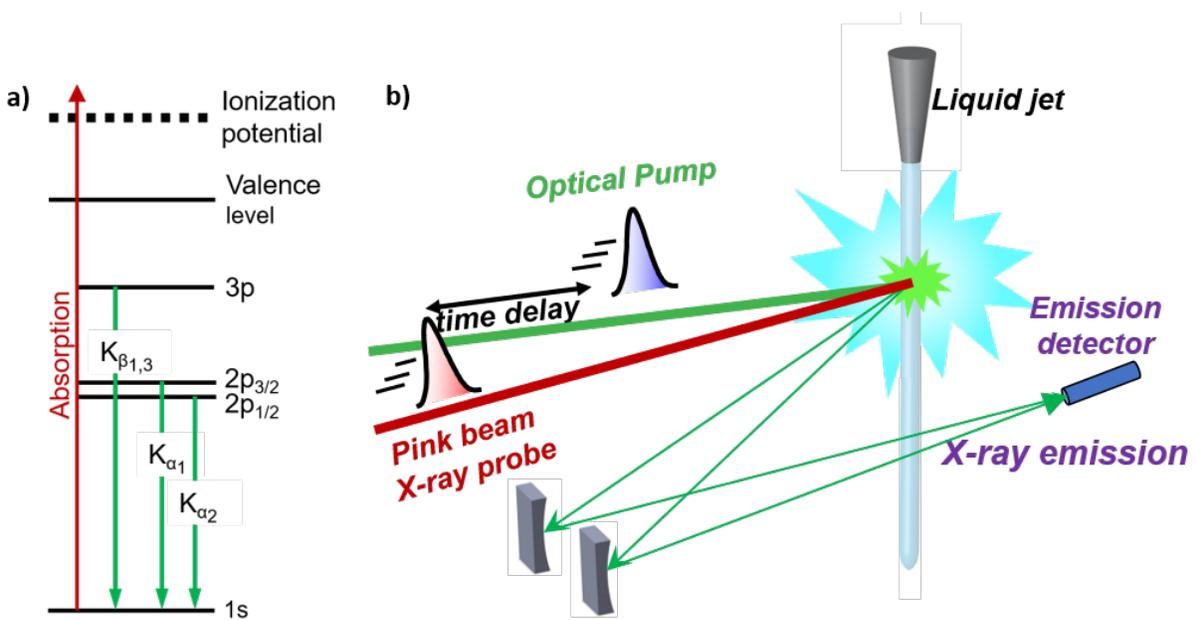

Figure 2: a) Energy level diagram showing the origin of the $K_\alpha$ and $K_\beta$ fluorescence after creation of a hole in the 1s (K) shell. The $K_{\alpha 1}$ and $K_{\alpha 2}$ lines originate from the splitting of the 2p orbital ($2p_{1/2}$ and $2p_{3/2}$), whereas for $K_\beta$ these lines are degenerate, resulting in the line labelled $K_{\beta 1,3}$. b) Experimental setup for the time-resolved X-ray emission spectroscopy measurements at the XFEL. A von Hamos geometry was used for these measurements.

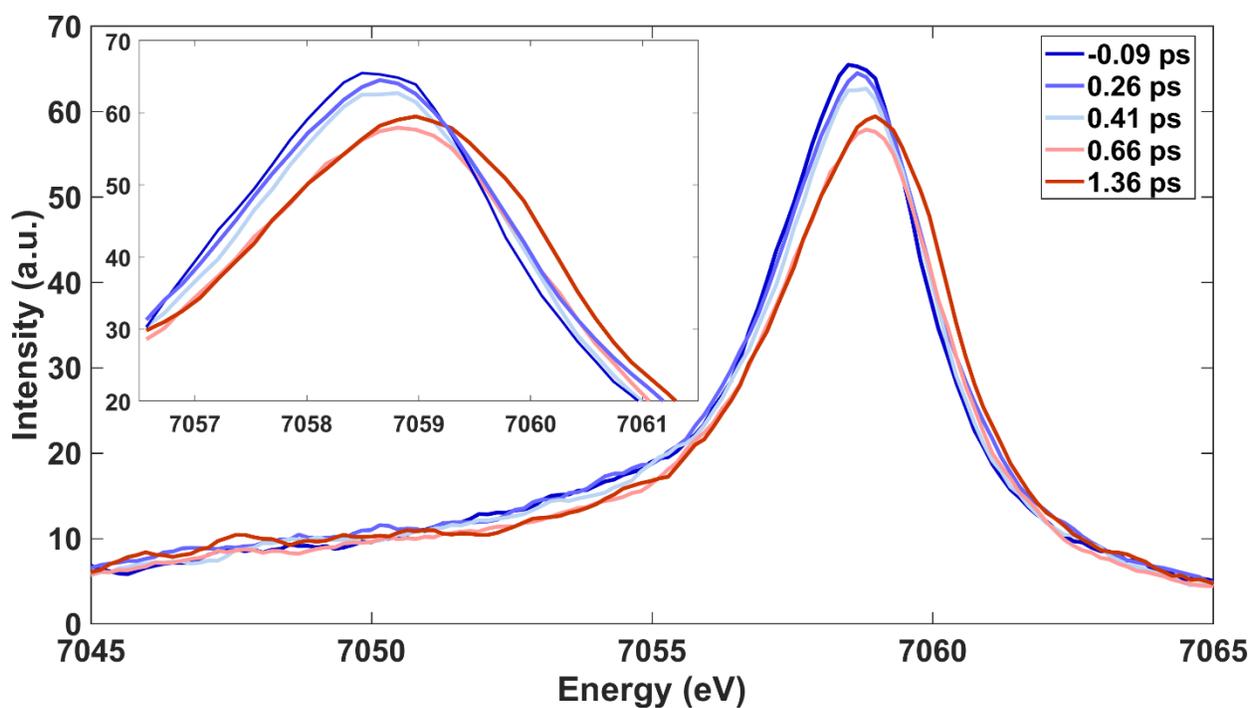

Figure 3: Laser-off (unpumped) and Laser-on (pumped) $K_\beta$ XES spectra of MbNO at different time delays between -0.09 and 1.36 ps (from blue to red) showing a blue shift of the $K_{\beta1,3}$ line and an intensity decrease. The inset zooms the region of the maximum of $K_{\beta_{1,3}} XES$ line at different time delays between -0.09 and 1.36 ps (from blue to red) showing peak shifts smaller than the energy resolution (~0.5 eV) of our experiment.

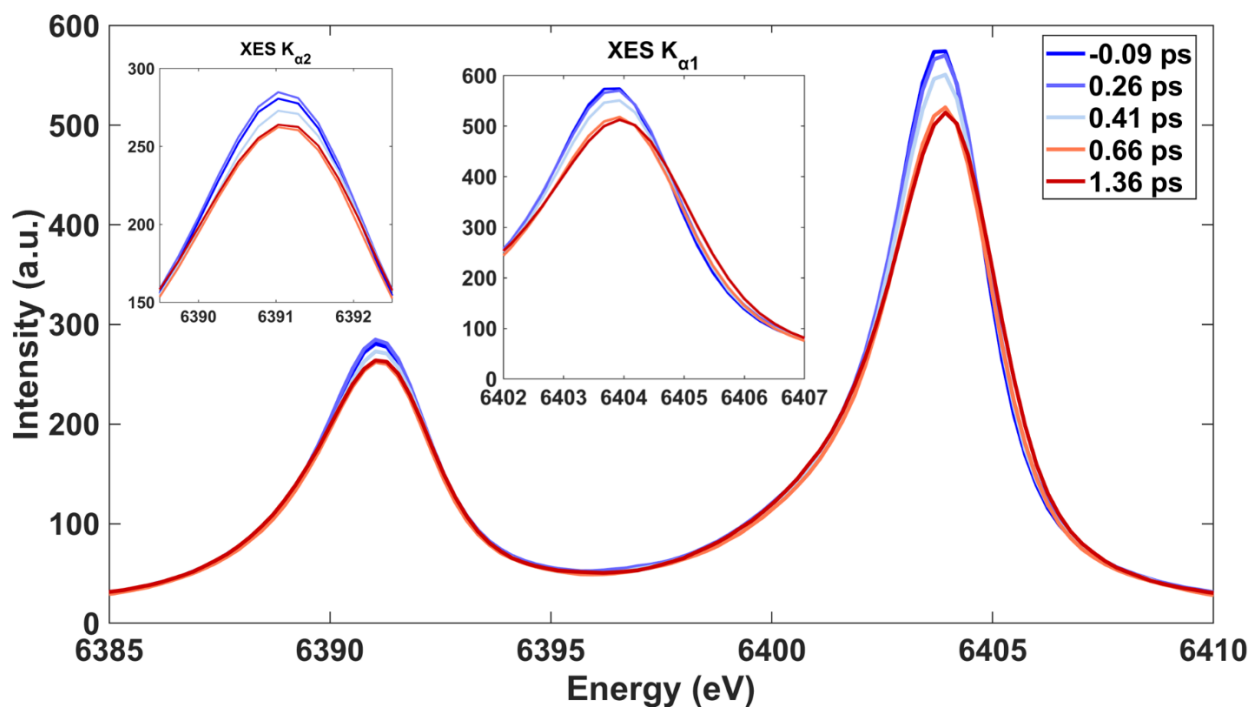

Figure 4: Laser-on (pumped) $K_\alpha$ XES spectra of MbNO at different time delays between -0.09 and 1.36 ps (from blue to red) showing an intensity decrease for $K_{\alpha 2}$ over the first 1.2 ps after excitation and an intensity decrease and change in peak width for $K_{\alpha 1}$. Also, the visibility of changes in peak width (σ) smaller than the energy resolution (~0.6 eV) can be well observed. The insets zoom into the peaks of the $K_{\alpha 2}$ and $K_{\alpha 1}$ laser-on (pumped) XES spectra of MbNO at different time delays between -0.09 and 1.36 ps (from blue to red) showing an intensity decrease for $K_{\alpha 2}$ over the first 1.2 ps after excitation and an intensity decrease and change in peak width for $K_{\alpha 1}$. The changes in peak width (σ) smaller than the energy resolution (~0.6 eV) can be well observed.

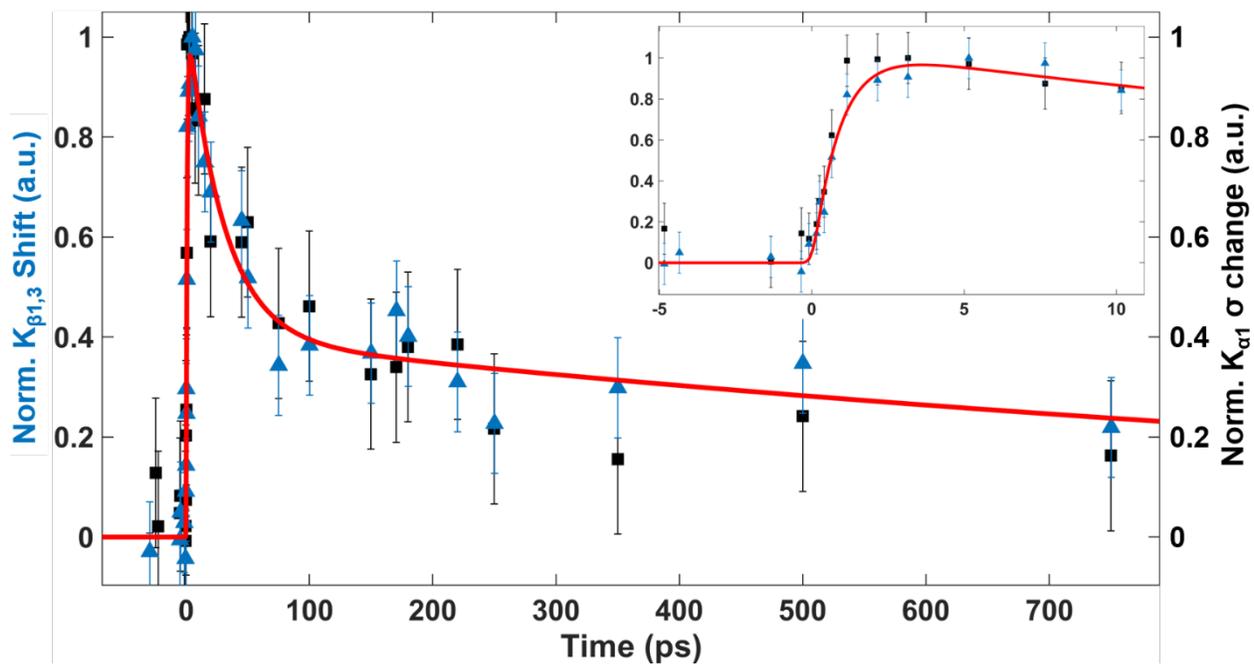

Figure 5: Temporal evolution of the relative $K_{\beta_{1,3}}$ shift for MbNO (blue triangles, centroid obtained from Gaussian fit, max shift is ~0.45 eV, see § S6) and of the $K_{\alpha_1}$ peak width changes for MbNO (black squares, normalized σ obtained from a Gaussian fit, maximum change in σ is ~0.3). The inset shows the temporal evolution during the first 10 ps of both signals. Details are explained in § S6.

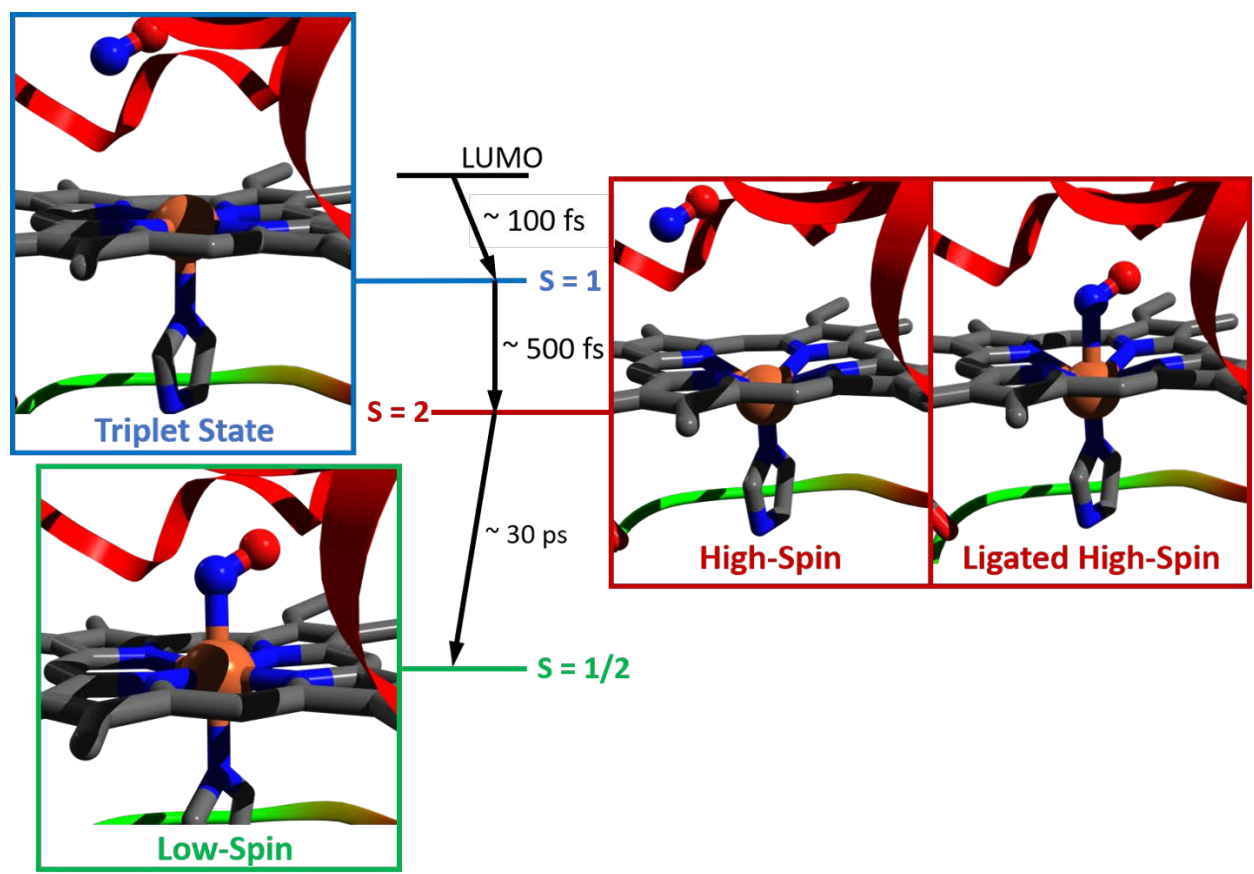

Figure 6: Schematic representation of the spin cross-over and reverse spin cross-over characterizing the photocycle of MbNO. The low-spin ground state MbNO ($S = ½$) haem undergoes prompt dissociation of NO upon $\pi-\pi^*$ excitation. The porphyrin Q-state decays to a triplet state ($S = 1$) of the pentacoordinated deoxyMb haem in $< 100$ fs.[47] Further relaxation to the quintet state occurs in ~500±250 fs. The entire process occurs in ~800 fs. Upon recombination of NO to deoxyMb leading to a hexacoordinated HS domed haem, relaxation back to the LS planar ground state occurs in ~30 ps. The LS structure is PDB entry 2FRJ, the sizes of the Fe atom, the NO molecule and the doming are exaggerated in order to highlight the key changes in the photocycle. The actual structural differences between MbNO and deoxyMb are shown in Figure S1. Fe = orange, N = blue, O = red and C = grey.